# Advances in Secondary Ion Mass Spectrometry for N-Doped Niobium

J. Angle[1], Ari Palczewski[2], C.E. Reece[2], F.A. Stevie[3], M.J. Kelley[1,2]

[1]Virginia Polytechnic Institute and State University, Blacksburg, VA, USA

[2]Thomas Jefferson National Accelerator Facility, Newport News, VA, USA

[3]Analytical Instrumentation Facility, North Carolina State University, Raleigh, NC,

## ABSTRACT

Accurate SIMS measurement of nitrogen in niobium relies on the use of closely equivalent standards, made by ion implantation, to convert nitrogen signal intensity to nitrogen content by determination of relative sensitivity factors (RSF). Accurate RSF values for ppm-range nitrogen contents are increasingly critical, as more precision is sought in processes for next-generation superconducting radio-frequency (SRF) accelerator cavities. Factors influencing RSF value measurements were investigated with the aim of reliably attaining better than 10% accuracy in N concentrations at various depths into the bulk. This has been accomplished for materials typical of SRF cavities at the cost of great attention to all aspects.

## INTRODUCTION

Particle accelerators are an important family of research instruments which use a microwave electric field in a resonant cavity to propel charged particles to GeV-range energies. The leading machines rely on superconducting radiofrequency (SRF) technology operating at 2 K to achieve optimal performance [1]. X-ray free electron lasers are a current example. The first generation machines European X-ray Free Electron Laser (E-XFEL) and the Linac Coherent Light Source (LCLS) showed the potential to reveal significant new information about bio-active proteins *in situ*. The prospective science that could be accomplished motivated a decision to build a next generation machine – LCLS-II- currently under construction at SLAC National Accelerator Laboratory [2].

Fortunately, the discovery of nitrogen doping at Fermi National Accelerator Laboratory (FNAL) occurred during the design phase of the LCLS-II[3]. It allows for cavities to operate at a higher efficiency (increased Qo) resulting in reduced initial and operating costs[4]. This led to the adoption of nitrogen doping technology for LCLS II and a scale-up program was initiated. Despite favorable results in single cell cavities, reproducing them in nine-cell cavities encountered difficulty[2]. An early area of concern was the possibility of varying nitrogen levels in vendor-supplied materials. An analytical method was needed that could quantitatively determine nitrogen contents to well below 100 ppm. The semiconductor industry faces similar problems with doping and has shown that dynamic SIMS could be developed for this purpose.

SIMS quantitation can be achieved by comparing the dopant signal intensities of both an experimental and a known reference material. The reference material matches the experimental matrix with a dopant concentration near that of the experimental. The relationship of the measured signal intensity to the known dopant content (relative sensitivity factor – RSF) affords a path to quantitation when the content is unknown. Using this method, SIMS performed by JLab and Virginia Tech allowed for investigation of vendor supplied LCLS-II material. A major finding was that the as-received nitrogen content of all the vendor- supplied material was within specification[5].

While the design of LCLS II was underway, additional studies indicated that shorter x-ray wavelengths would enable further unique experiments. Such x-ray wavelengths could be obtained from a higher energy electron beam from the accelerator, obtained in turn from higher performance from the SRF cavities. Accordingly, an upgrade program was launched – LCLS II HE[6]. An experiment-based method for modeling cavity performance was needed, requiring a more precise SIMS method. Previous quantitation found that the RSF measurements could vary by as much as 50%; a value far more than the <10% target [5] [7]. Our objective in this study is to improve SIMS quantitation of nitrogen in niobium. The issue is error in RSF values calculated from reference materials. Here we report investigation of the factors which cause such RSF uncertainty and demonstrate a mitigation strategy.

**SIMS**

The concentration of dopants, typically in the ppm range, affects the electrical properties of a semiconductor. Accurately knowing their concentrations is important to achieving the desired properties. For decades SIMS has been an invaluable semiconductor industry tool to accurately quantify ppm levels of dopants[8]. Niobium SRF materials offer several complications. The specimens are typically polycrystalline and may have notable surface topography due to (e.g.) preferential etching of the grains during surface treatments. Grain sizes range from ~ 50 µm typical of vendor-supplied sheet stock to more than 200 µm after high temperature annealing. The relationship of feature size to primary beam raster size must be considered. Furthermore, commercial and traceable reference materials do not exist for niobium materials and must be fabricated by ion implantation. In contrast to implanted silicon materials, implanted niobium standards cannot be cross-checked by RBS due to poor RBS detection limits for light elements in heavy element matrices. These are among the challenges that need to be addressed to achieve accurate and precise measurement of nitrogen content in niobium matrices.

To properly quantify the concentrations of dopants, the instrument must first be calibrated with standards prepared by implanting known doses of nitrogen in a niobium matrix. Analysis of the implant standards enables determination of the relative sensitivity factor used to convert detected nitrogen signal intensity to nitrogen concentration[9]. The formula is shown below

$$RSF = (\Phi C I_m t)/(z I_s)$$

Where $\Phi$ is the implanted dose of the target species (here, N), C the number of data collection cycles, $I_m$ is the intensity of the matrix species, t time per cycle, z is the crater depth, and $I_s$ is the summation of the intensities from the reference species. The crater is idealized as a constant depth with a flat bottom in a flat specimen surface.

A critical component of the RSF calculation is the crater depth measurement. The quality of the instrument tune can affect the crater shape and is a property well understood by SIMS analysts. Flat crater bottoms are required to precisely measure sputter rate. Uncertainty in the sputter crater depth increases the uncertainty of the calculated RSF value. Another factor which can affect the crater depth determination is the surface roughness of the implant standard. Further, different Nb grain orientations may have different sputter rates, which motivates quantitative profiling of single grains.

A further topographical effect on the apparent secondary ion yield has been studied in the context of the observed concentration of oxygen isotopes in geological samples[10-13]. The difference in sample topography causes changes in the surface electrostatic field which alters the trajectory of the secondary ions as they enter the column. Mitigation is possible by adjusting the dynamic transfer contrast optics prior to data collection to center secondary ions as they approach the entrance slit[14, 15]. Our previously reported SIMS experiments, which yielded 50% uncertainty of the RSF values, were performed without the dynamic transfer optics correction. Here we report the effectiveness of use of this secondary beam tuning correction for SRF materials.

**Experimental**

The SIMS measurements were performed on a CAMECA 7f Geo magnetic sector SIMS instrument. A $Cs^+$ primary ion beam with an accelerating voltage of 5 kV was rastered over the surface with a current initially starting at 100 nA. The beam current was later lowered to 25 nA to increase the quality of the beam tune by reducing its spot size. The beam was set to raster over a 150 μm x 150 μm area and data were collected from a 63 μm x 63 μm area in the center of the raster. $^{12}C^-$, $^{16}O^-$, $^{93}NbN^-$ secondary ions were detected in conjunction with a $^{93}Nb^-$ reference signal. $NbN^-$ was monitored as the indicator for N, since the secondary ion yield for $N^-$ is negligible. Experiments performed included the analysis of polycrystalline, single crystal and bicrystal samples, and an investigation of the sample holder geometry's effect on RSF. Beam centering by alignment of the dynamic transfer contrast optics (DTCA-X and DTCA-Y) was used only during experiments where mentioned.

Following the SIMS data collection, a KLA-Tencor Alpha Step 500 stylus profilometer was used to determine crater depths. The stylus was scanned in both axes of the SIMS crater. The stylus was scanned at a rate of 50 μm/s to acquire a 500 μm profile. Results from both axes were averaged to yield a net value. CAMECA WinCurve data analysis software was used to determine the RSF and subsequently convert the depth profile data from signal intensity to concentration.

Crystal orientation was determined by electron backscatter diffraction (EBSD). An FEI NanoLab SEM/FIB outfitted with an EDAX TSL EBSD camera was used to acquire the orientation image maps (OIM). The samples were mounted with a specimen pre-tilt of 70°, at which the OIM maps were acquired with an electron beam voltage of 30 kV and beam current of 21 nA.

**Samples**

To ensure samples fit the CAMECA 7f sample holder with maximum efficiency, 10 x 6 mm coupons were cut by electrical discharge machining (EDM). Both polycrystalline and large bicrystal or single crystal specimens were prepared. The specimens were polished by a series of steps including buffered chemical polishing (BCP), nanopolishing (NP), and electropolishing (EP) [16-18]. Experimental samples were nitrogen doped at the Thomas Jefferson National Accelerator Lab (JLab) using recipes in Table 1. Following the doping process, an additional 5 µm EP was performed to remove the surface nitrides. The samples designated as implantation standards were sent to Leonard Kroko, Inc. where they were implanted with carbon, oxygen, and nitrogen. Early implant standards such as NL 133 were implanted with all three target species. However, to avoid possible mass interferences between $NbN^-$ and $NbCH_2^-$ the implantation of carbon was dropped for L78, L79, and U52. Shown in Figure 1 is a representative depth profile for the implantation standards.

**Table 1.** Process history of implant standards and experimental samples.

| Sample | Type of Sample |
|---|---|
| **NL 133** | Implant Standard – 30 µm EP, pre-annealed at 900C before cutting |
| | $2x10^{15}$ atoms/cm$^2$ of carbon at 135 keV |
| | $2x10^{15}$ atoms/cm$^2$ of oxygen, at 180 keV |
| | $2x10^{15}$ atoms/cm$^2$ of nitrogen, at 160 keV |
| **NL 134** | Implant Standard – 30 µm EP, pre-annealed at 900C before cutting |
| | $2x10^{15}$ atoms/cm$^2$ of carbon at 135 keV |
| | $2x10^{15}$ atoms/cm$^2$ of oxygen, at 180 keV |
| | $2x10^{15}$ atoms/cm$^2$ of nitrogen, at 160 keV |
| **NL 136** | Implant Standard – 30 µm EP, pre-annealed at 900C before cutting |
| | $2x10^{15}$ atoms/cm$^2$ of carbon at 135 keV |
| | $2x10^{15}$ atoms/cm$^2$ of oxygen, at 180 keV |
| | $2x10^{15}$ atoms/cm$^2$ of nitrogen, at 160 keV |
| **NL 115** | N-Doped 3N60 inside the beam tube of EZ SSC-05 – 30 µm EP, pre-annealed at 900C before cutting, after doping HF soak for 9 hours total, 5 µm EP |
| **NL140** | N-Doped 2N0 inside KEK04, pre-annealed at 900C before cutting, |
| **NL 147** | N-Doped 3N60 inside HE-353, no pre-annealing performed. No EP following doping |
| **L 78** | Niobium Implant Standard – 30 µm BCP |
| | $2x10^{15}$ atoms/cm$^2$ of oxygen, at 180 keV |
| | $2x10^{15}$ atoms/cm$^2$ of nitrogen, at 160 keV |
| **L79** | Niobium Implant Standard – 30 µm BCP |
| | $2x10^{15}$ atoms/cm$^2$ of oxygen, at 180 keV |
| | $2x10^{15}$ atoms/cm$^2$ of nitrogen, at 160 keV |
| **U 52** | Niobium Implant Standard – 50 µm EP nanopolish |
| | $2x10^{15}$ atoms/cm$^2$ of oxygen, at 180 keV |
| | $2x10^{15}$ atoms/cm$^2$ of nitrogen, at 160 keV |
| **Si# 2293** | Silicon Implant Standard |

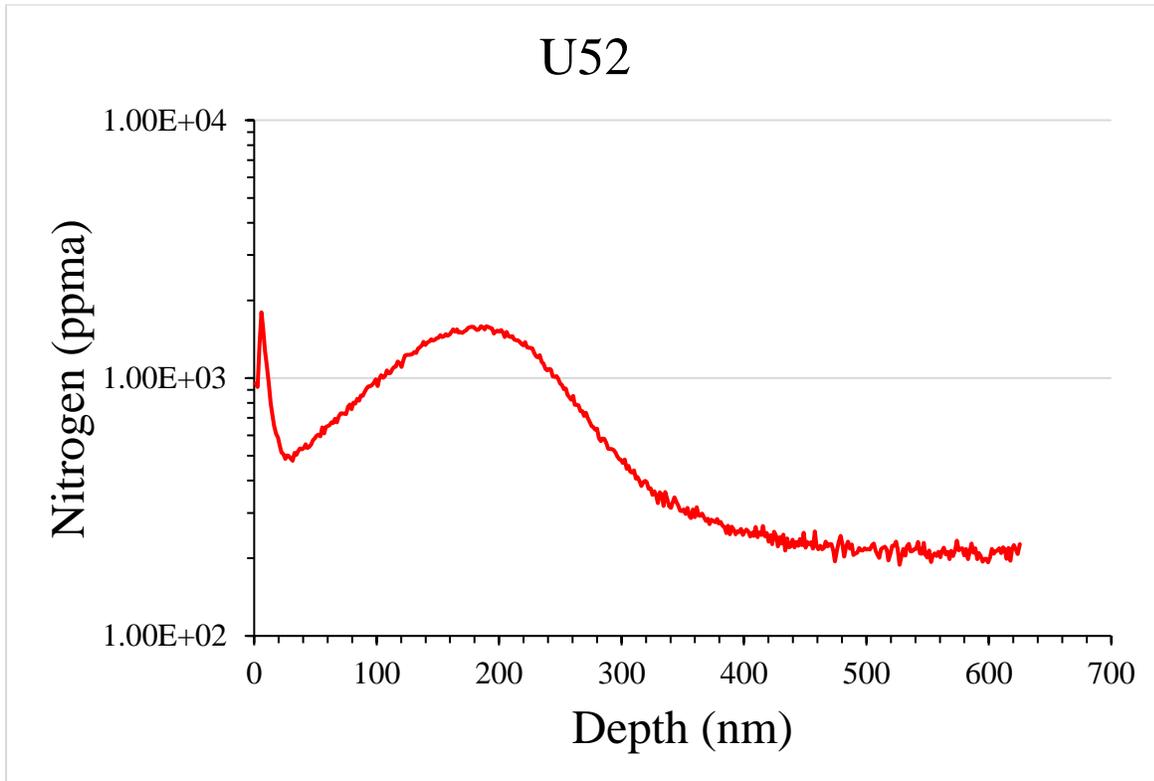

**Figure 1.** Representative NbN$^-$ depth profile of an implantation standard. Depicted above shows implant standard U52 which contains 2x10$^{15}$ atoms/cm$^2$ of nitrogen dosed at 160 keV into a niobium matrix.

**RESULTS AND DISCUSSION**

**A. Polycrystalline Materials**

Our previous SIMS measurements of polycrystalline N-doped specimens reported an apparent concentration difference between grains of different crystallographic orientation[5]. Specifically, a large difference in ppma was observed between [001] and [111] orientations on the same sample[19]. In addition, the sputter rates varied from grain to grain. Therefore, investigation to understand the cause became necessary. To determine the effect of grain orientation on the instrument calibration or nitrogen uptake, EBSD was performed on the specimens to identify the surface normal grain orientations.

As a first step, orientation image maps (OIM) were acquired from locations covering more than several square millimeters on the surface of a nitrogen-implanted polycrystalline specimen (NL 133) and combined to obtain a single OIM of the sample surface (Figure 2 a.) SIMS data were collected at the locations indicated by the squares. An RSF was calculated for each (Figure 2 b.) However, no correlation between grain orientation and RSF was observed. Further, the RSF was observed to vary substantially, yielding a 40% RSD (relative standard deviation). Replicate implant standards were analyzed and found to vary 16% and 14% for NL 134 and 136 respectively (Figure 3).

Though samples NL 133, NL 134 and NL 136 were electropolished as a final preparation step, grain height variation was noted. Electropolishing when performed under optimal conditions will yield a sub-100 nm finish[16, 20-22]. Differential surface removal has been observed in previous studies and achieving an ideal surface is non-trivial[23]. Continual improvement is the subject of further investigation.

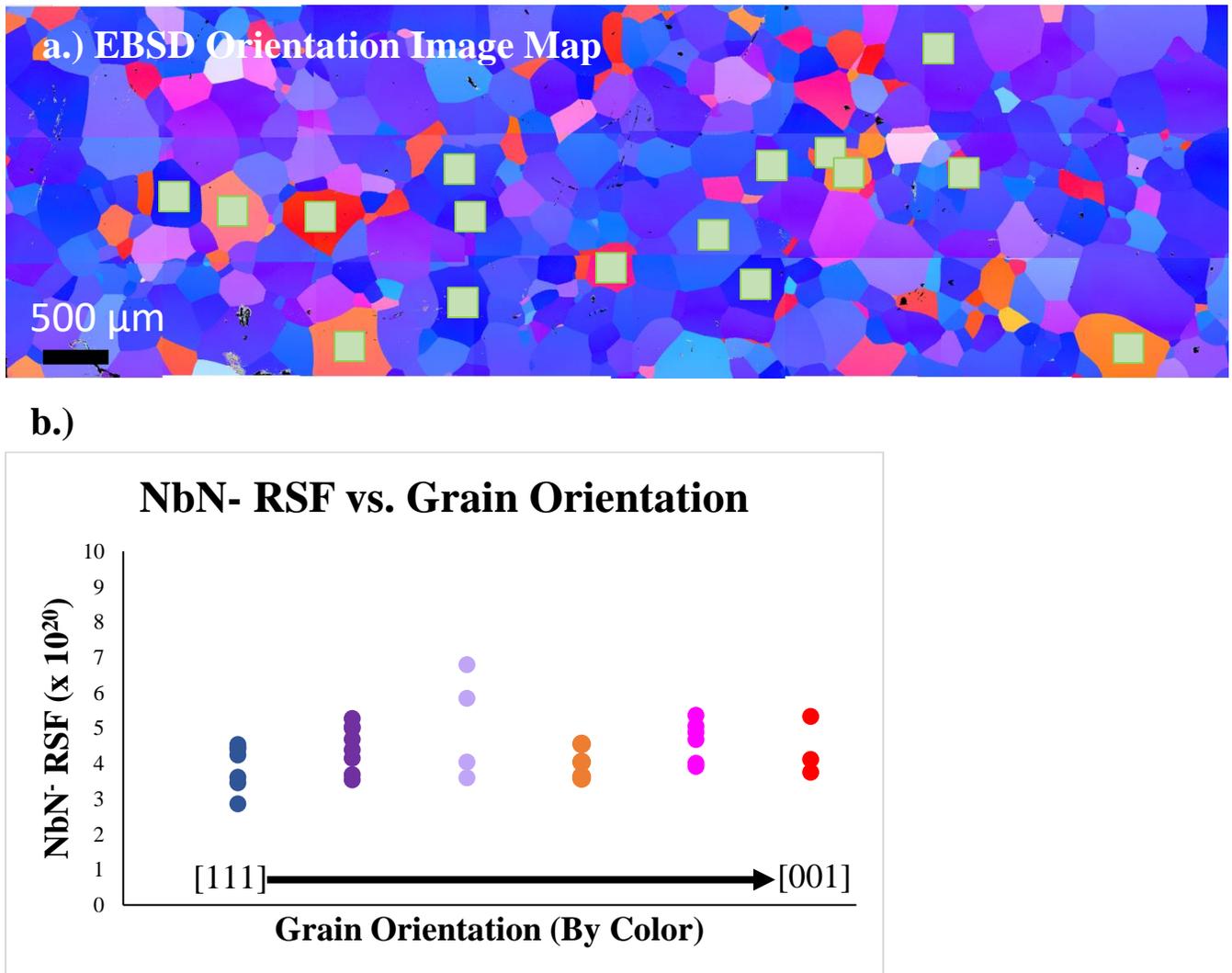

**Figure 2.** a) EBSD OIM map of the NL 133 Implant Standard. Locations where SIMS analyses were performed are noted. b) NbN⁻ RSF was calculated for each data point and plotted as a function of grain orientation. No discernable correlation was found.

### B. Single/Bicrystal Implant Standards

To gain further insight into the effect of grain orientation on RSF, single and bicrystal samples were analyzed. Due the pristine and flat nature of the single crystals, reproducible and consistent RSF could be established when compared to polycrystalline samples (Figure 3). To better understand the effect of sample positioning on the RSF, single crystal implant standard U52 was rotated in 90º increments in the sample holder. It was observed that the sputter rate changed as a result of sample rotation, but the RSF was unaffected. Though the analysis time and sputter depth are variables which affect the RSF calculation, this suggests the nitrogen and niobium intensities adjusted to normalize the change in sputter rate.

Implant standard U52 was found to have the most precise RSF values compared to the L78 and L79 large grain implant standards, as a result of U52 having a superior surface finish (Figure 4). Accurate crater depth determination is reliant on the crater bottom being flat and parallel to the sample's external surface. Poor instrument tuning or rough standard finishes introduce error into the crater depth determination which propagates to the RSF determination. Figure 5 shows a comparison of profilometry scans between L79 and U52.

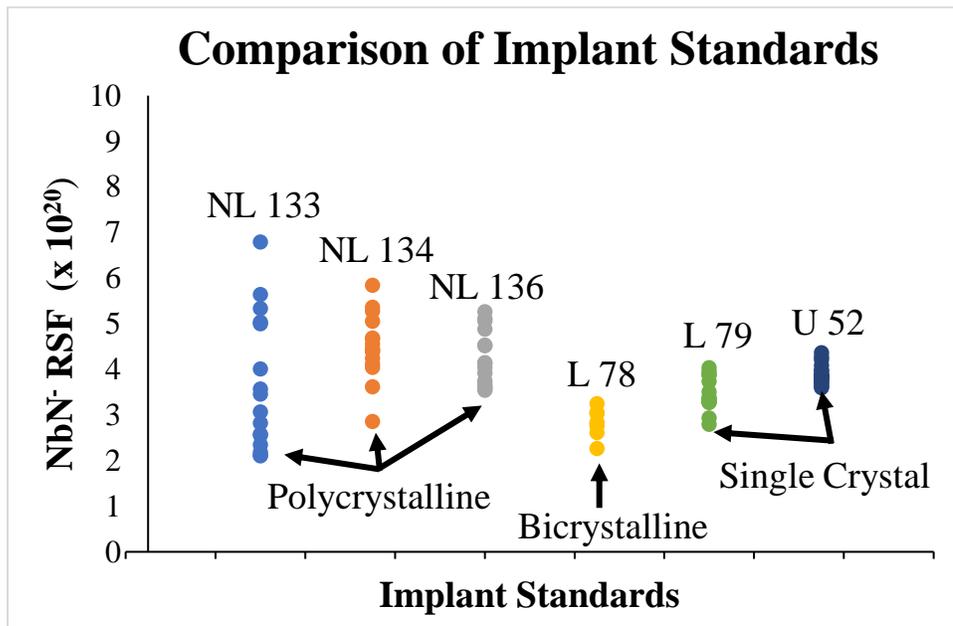

**Figure 3**. Plot depicting RSF values per analysis. Each column shows analyses run for a specified sample. Changing from polycrystalline standards to single crystals yielded more precise RSF values. Further increasing the surface quality to the smoothest finish (U52) increased the precision.

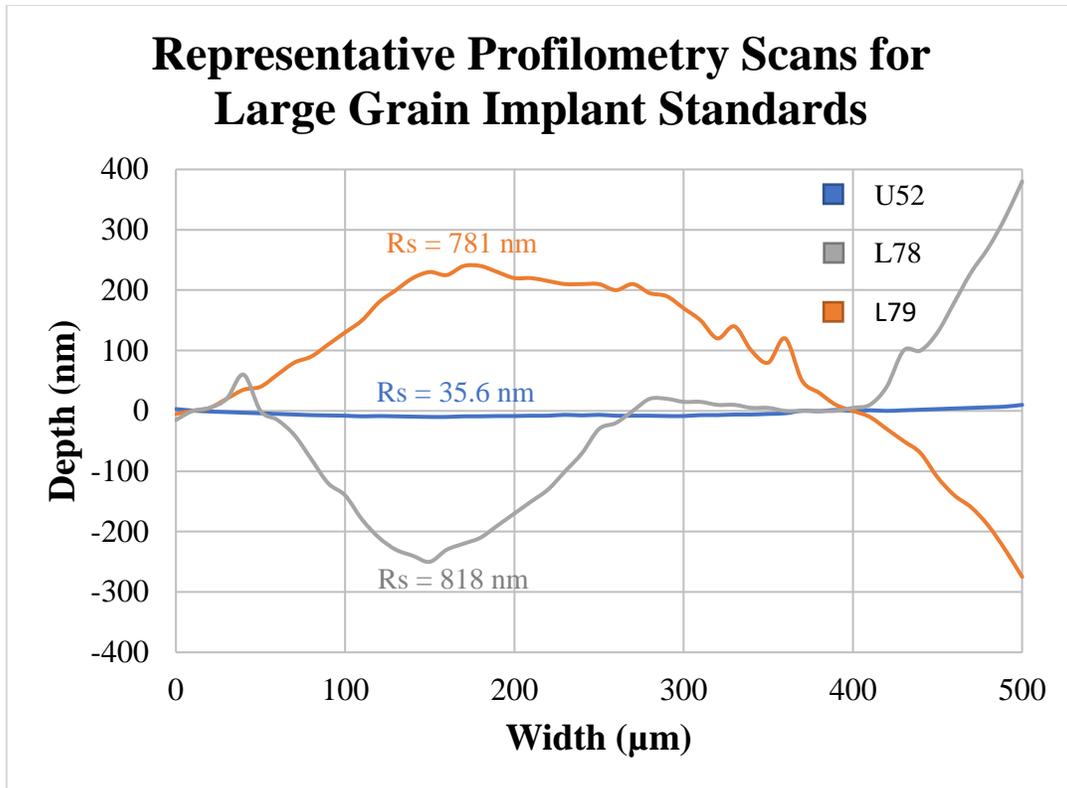

**Figure 4**. Representative profilometry scans of the single and bicrystal implant standards. The average surface roughness for each standard is annotated.

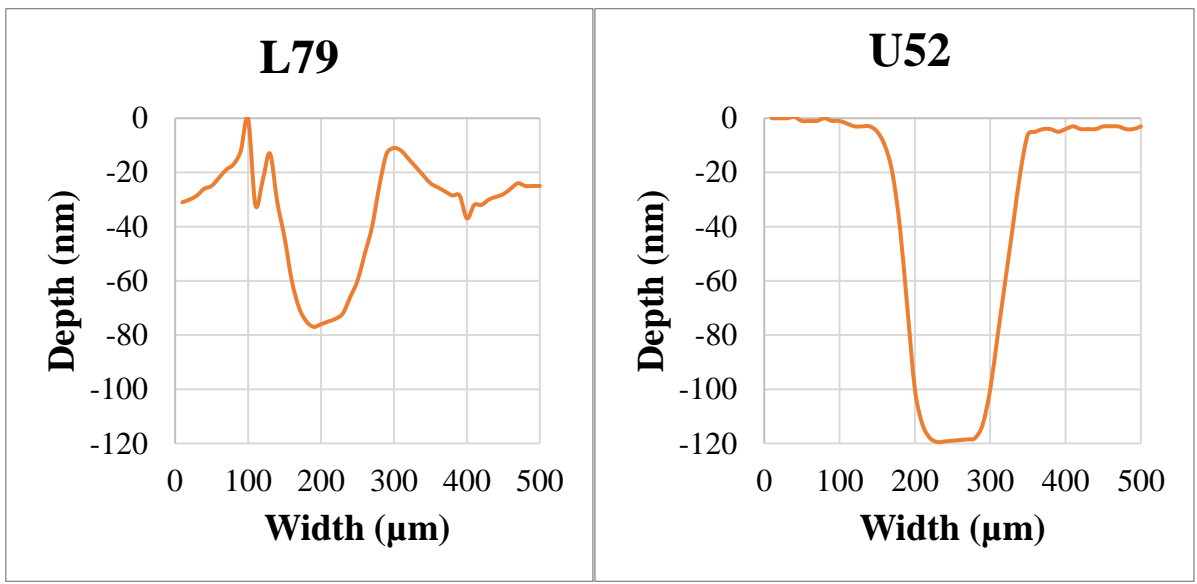

**Figure 5**. Comparison of profilomety scans for single crystal implant standards. The surface finish for U52 was observed to be much smoother which yields accurate crater depths and minimizes the propagation of error for RSF determination.

### C. CAMECA Sample Holder- DTCA effectiveness

It was observed that the U52 implant standard yielded consistent RSF values. However, in the experiment, the specimen was loaded into the sample holder in the same location and analyses were taken given the same sample loading conditions. Peres et al. (2011) observed that the topography effect on the reproducibility of SIMS analyses is not limited to the sample but includes the sample holder. As a result, the CAMECA 7f Geo specimen holder was investigated to uncover effects which may change the RSF. This holder is capable of loading 1-4 specimens approximately 5 mm x 5 mm loaded individually or to full capacity (Figure 6a). The specimens are held in place by springs that apply a flexural load to the thin faceplate of the sample holder (Figure 6a). To determine if the loading force variation may result in a change in RSF, SIMS analysis was performed on implant standard U52 in which the sample holder contained 2 specimens and repeated with 4 specimens. The results showed a clear change in RSF because of the loading force change (Figure 6 b,c). The impact of changing the loading force was observed by monitoring the niobium matrix signal. The results showed a signal change as a function of loading force (Figure 6 b). These findings suggest that the loading force can change the working distance of specimens, which can in turn cause a trajectory change of the secondary ion beam as it approaches the secondary column, resulting in signal change and alteration of the system calibration.

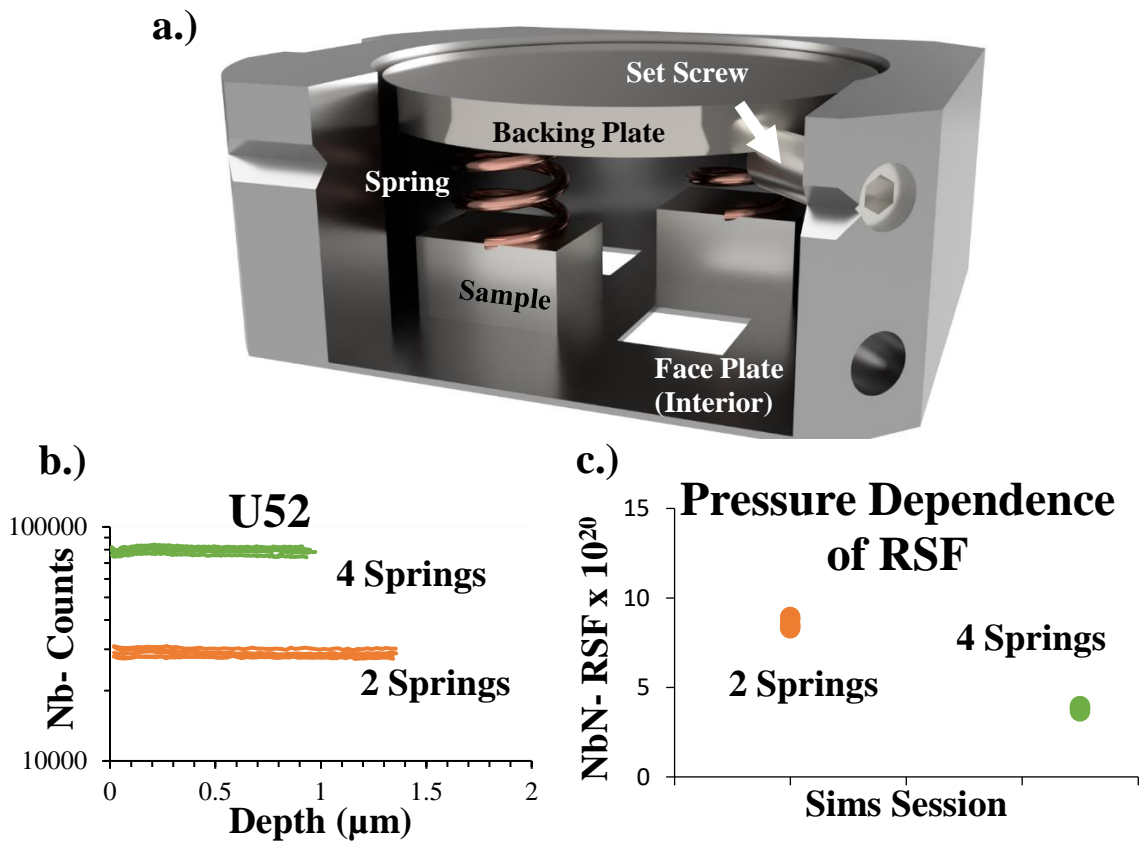

**Figure 6.** a) Depiction of a loaded CAMECA 7f Geo sample holder. Samples are placed into position and held in place by compressing copper springs between the sample and the backing plate. The force applied can cause deflection of the face plate resulting in an observable change in b) the niobium matrix signal and the c) RSF.

Topographical variations of the sample itself are expected in addition to working distance changes caused by sample loading. Therefore, to better understand how the RSF may change as a function of position on a sample holder, a single silicon implant standard denoted as Si-N 2293 was fractured and loaded into three of the four slots of the sample holder. The remaining spot was loaded with a different Si-N implant standard with the same dose but different implantation energy to determine reproducibility between implant standards. Figure 7 shows the SIMS analysis locations. A proposed method for mitigating these topographical effects is to center the beam by using the dynamic transfer contrast apertures (DTCA). The DTCA can be adjusted in two dimensions in which the aperture is scanned in both the X and Y axis. The beam centering function scans on both lateral dimensions of the beam to correct the trajectory of the secondary beam as it enters the column. To compare the effectiveness of using this method, the experiment was performed with and without use of the DTCA. The results (Table 2) show that without the beam centering correction, the RSF varied 47% by only varying the analysis location. By using the beam centering parameter and adjusting the DTCA prior to each analysis the RSD was reduced to 1.5%. Points A and B were dosed the same as D through H but were from a different parent sample. Use of the DTCA revealed that the derived RSF was inherently different. This suggests that although implant standards may be nominally dosed the same, some variation may arise that translates into calibration error. Unlike many of the standards used in the semiconductor field, RBS cannot be used as a quality control for nitrogen implanted into niobium. This issue is noted, and the mitigation will be discussed in a future report. Without a secondary method, such as RBS, to check the dose, we are relying on the number provided by the implanter and that can be affected by mass interferences and by effects such as sample charging. One solution is to implant a piece of silicon at the same time and compare the result with prior implanted samples.

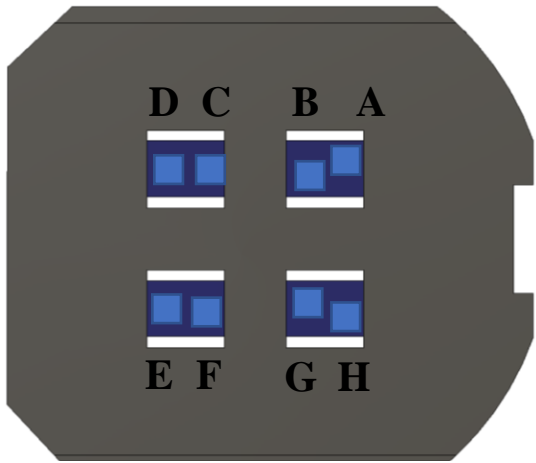

**Figure 7.** A map of the SIMS analysis showing location of RSF values as a function of sample position. Corresponding data are located in Table 2.

**Table 2.** Data from the RSF location mapping. Locations correspond with Figure 7. The analysis was performed with and without the DTCA beam centering function. The results clearly show an improvement in RSF reproducibility as well as matrix intensity. Note: as the beam centering function increased the overall counts of the analysis, the matrix signal was changed from- $^{28}Si_3$ to $^{28}Si_4$ to prevent damage to the electron multiplier.

| Si N 2293 | Without Beam Centering | | | With Beam Centering | | |
|---|---|---|---|---|---|---|
| | Depth AVG (nm) | $^{28}Si_3$ Intensity | RSF e20 | Depth AVG (nm) | $^{28}Si_4$ Intensity | RSF e20 |
| | | | | 642 | | 0.772 |
| A | 879 | 4.76E+05 | 8.11 | 801 | 3.78E+05 | 0.75 |
| B | 875 | 5.66E+05 | 8.7 | 770 | 3.22E+05 | 0.747 |
| Average | 877 | 5.21E+05 | 8.405 | 738 | 3.50E+05 | 0.75633 |
| StDEV | 2.8 | 6.36E+04 | 0.4 | 84 | 3.96E+04 | 0.01 |
| %RSD | **0.3%** | **12.2%** | **5.0%** | **11.4%** | **11.3%** | **1.8%** |
| C | 876 | 9.84E+05 | 17.9 | 771.6 | 3.35E+05 | 1.34 |
| D | 875 | 4.12E+04 | 5.75 | 748.3 | 3.32E+05 | 1.33 |
| E | 870 | 4.76E+04 | 6.42 | 753.4 | 3.44E+05 | 1.38 |
| F | 863 | 2.16E+05 | 11.8 | 761.9 | 3.49E+05 | 1.38 |
| G | 865 | 9.27E+04 | 8.66 | 751.8 | 3.31E+05 | 1.36 |
| H | 859 | 8.23E+05 | 16.6 | 738.5 | 3.47E+05 | 1.37 |
| Average | 868 | 3.67E+05 | 11.2 | 754 | 3.40E+05 | 1.36 |
| StDEV | 6.8 | 4.23E+05 | 5.2 | 11 | 7.94E+03 | 0.02 |
| %RSD | **0.8%** | **115.1%** | **46.2%** | **1.5%** | **2.3%** | **1.5%** |

### D. Practical Applications- DTCA

The implant standards previously tested were reanalyzed with the beam centering function. NL133, for which an RSF variation of 40% had been measured, yielded the most notable change in precision as the uncertainty was reduced to 9.7%. Reduction in uncertainties was observed for all other implants with the values noted in Figure 8. As was the case in the previous experiments, the RSF precision was superior for the single crystal and bicrystal standards.

Previous SIMS analysis of N-doped samples appeared to show variation in nitrogen content as a function of grain orientation. However, at the time of analysis, effects of surface topography were not accounted for. A profilometry scan of a SIMS analyzed [111] grain adjacent to (001) grains showed that the grain height could vary as much as 1.5 µm (Figure 9), which is sufficient to alter the trajectory of the secondary ion beam. Additional SEM analysis performed on sample NL 115 showed the presence of surface features on [111] grains that were absent from [001] grains (Figure 10), If these are nano-nitrides as some have suggested, then the nitrogen content of the [111] grains surface may be higher than the [001] grains, not the opposite. Therefore, sample NL 115 was re-analyzed with DTCA beam centering to determine the effectiveness of this method on such a sample. All scans showed an increase in nitrogen content versus previously reported, which is apparently a result of correcting the secondary beam trajectory. Additionally, the [111] grains now reported nitrogen content higher than that of the [001] grains, which is consistent with the notion that the observed features are indeed nano-nitrides (Figure 11)."Nano-nitrides" were also observed on sample NL 140, therefore the experiment was repeated to determine if the observation could be reproduced. This sample was also observed to have higher nitrogen content in grains with a [111] normal surface orientation (Figure 12).

Reduction of method uncertainty was deemed essential to providing reliable and reproducible data to correlate SIMS data and cavity performance. The cavity co-doped with Sample NL 147 was found to perform outstandingly well in rf testing (Fig. 13). SIMS was performed on the sample without removal of the surface nitrides by EP. The data showed a sharp decline in nitrogen content over 1.5 micron from approximately 50% to ~ 1000 ppma (Fig. 14). A further modest reduction in the nitrogen content was also observed --- after this nitride layer was removed followed by a return to ~1300 ppma. Understanding in detail how this nitrogen content and its variation by doping process conditions contributes to the corresponding SRF cavity performance is a matter of great interest and ongoing analysis. These process conditions are not yet optimized.

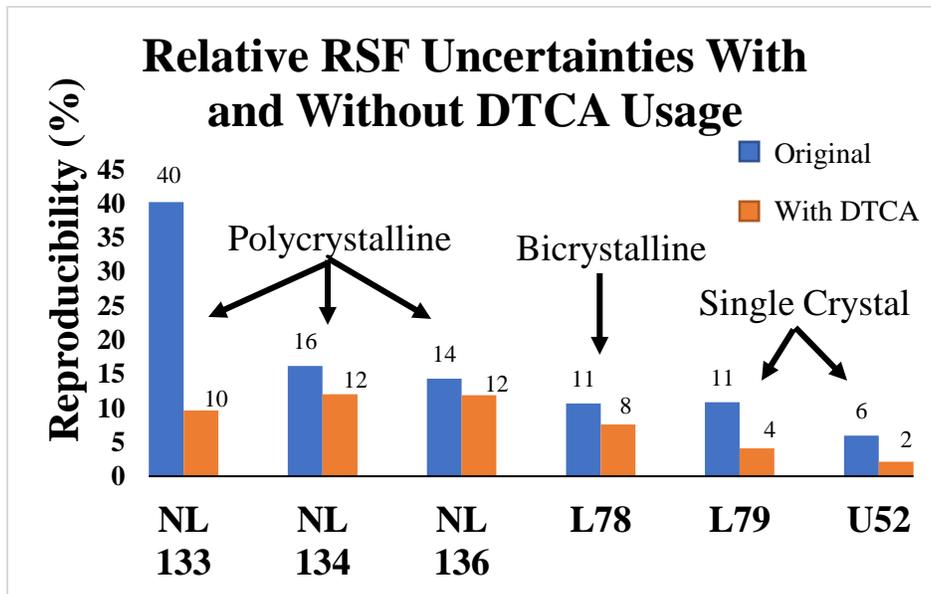

**Figure 8.** Comparison of RSF variation of implant standards. The plot shows that precision of RSF values improved for all tested implant standards when DTCA was used.

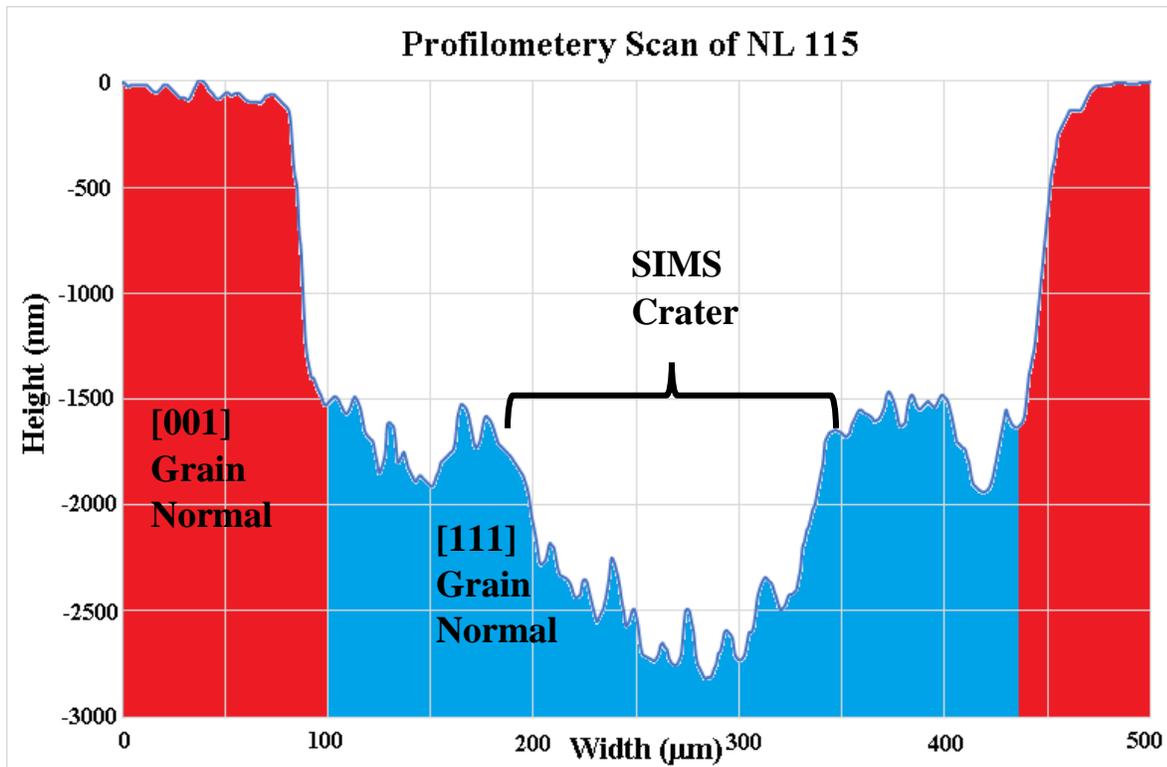

**Figure 9.** Profilometry scan of N-doped sample NL 115 in a region where a [111] grain is adjacent to [001] grains. The scan shows a height difference of roughly 1.5 µm between [111] and [001] grains. The [111] grains were observed to be especially rough due to the presence of "nano-nitrides" in the grain. SIMS was performed on this specific grain and the crater is noted.

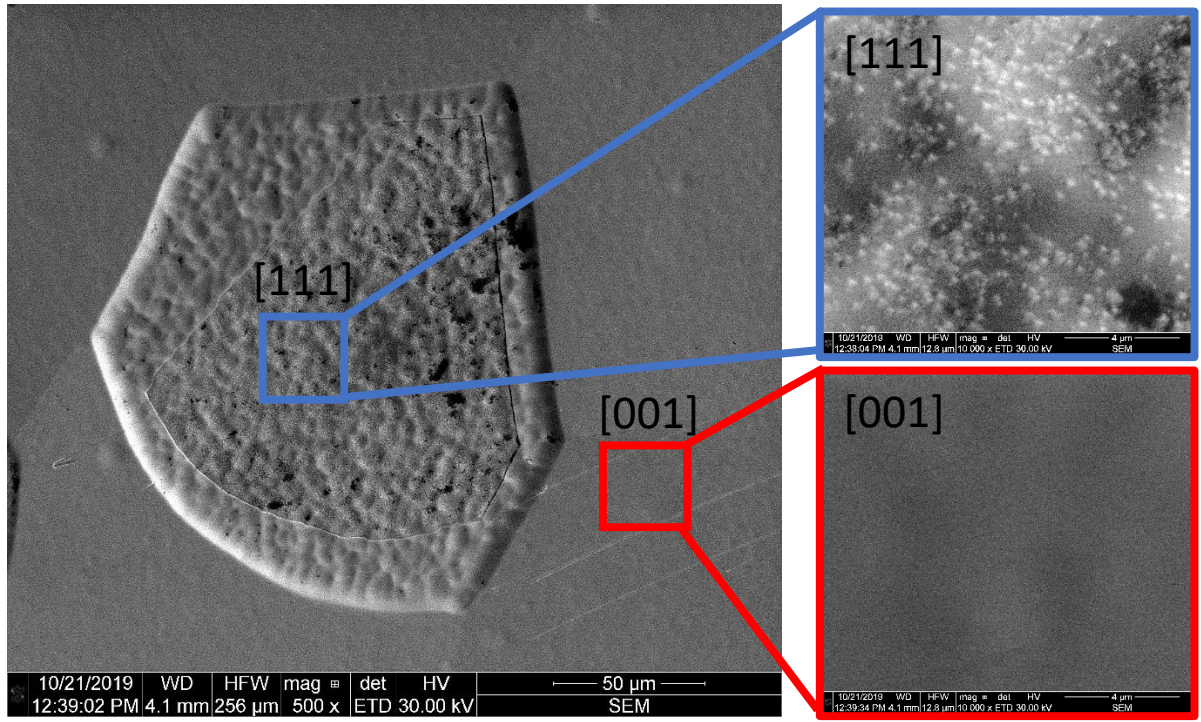

**Figure 10.** SEM image of N-Doped sample NL 115. The image shows particles hypothesized to be "nano-nitrides" present on near [111] grains but absent from [001] grains.

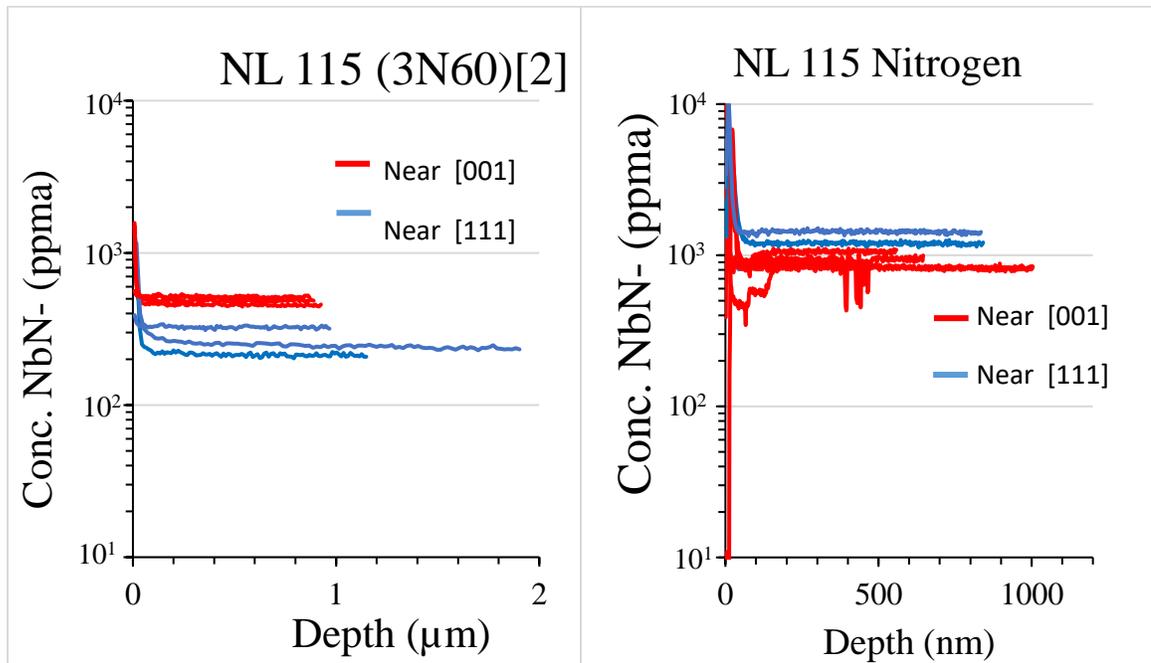

**Figure 11.** Comparison of SIMS data with and without beam centering. A.) previously reported results showed that the nitrogen content was unexpectedly lower in the [111] grains. B.) By centering the secondary beam with the DTCA aperture prior to the analysis the signal

intensity increased resulting in an apparent increase of nitrogen content. [111] grains now show a higher nitrogen content, consistent with the roughness observed on the [111] grains being nano-nitrides.

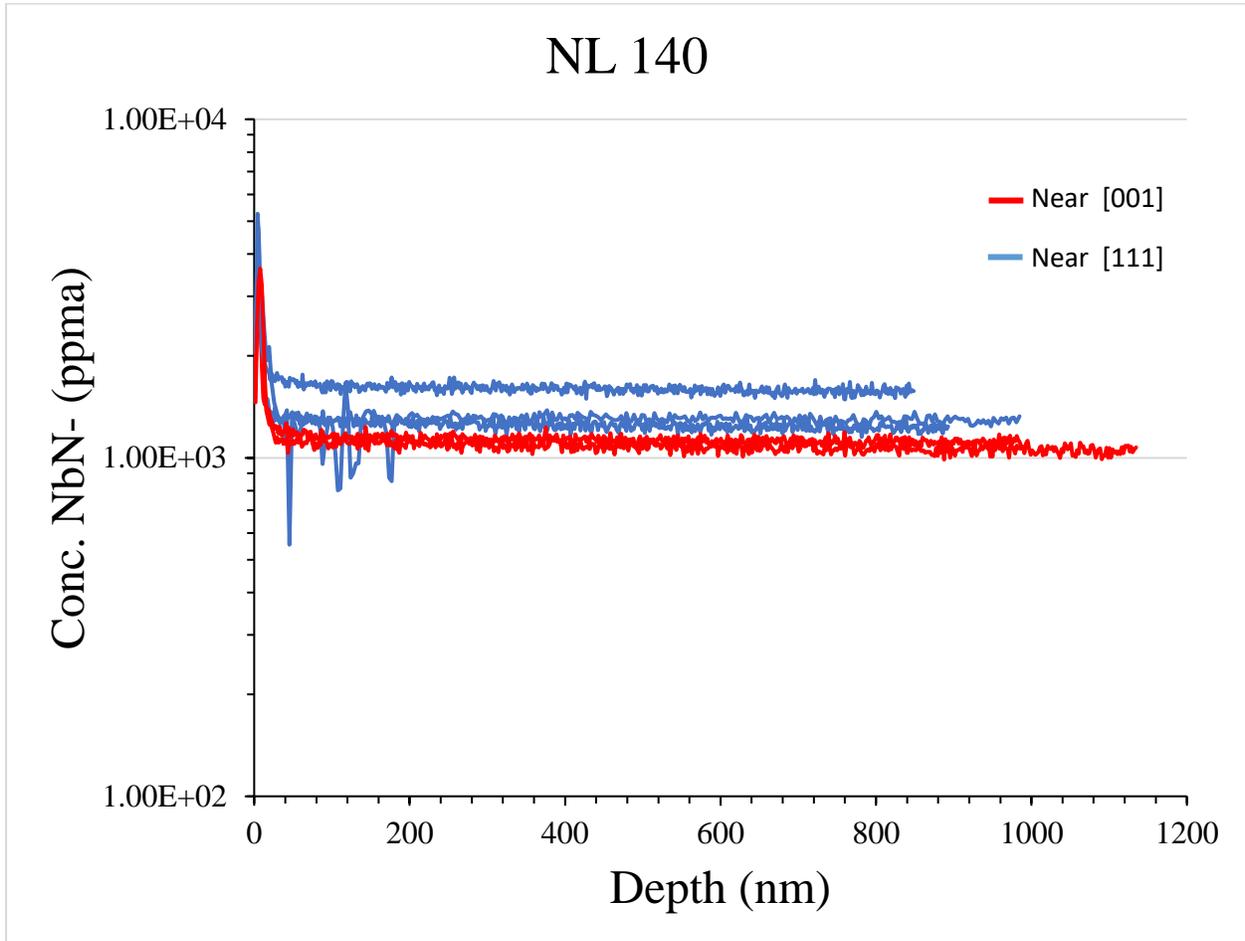

**Figure 12.** SIMS data for sample NL 140, an N-Doped sample with small particles observed on the surface. As was the case for sample NL 115, the NbN⁻ concentration was higher for near [111] grains.

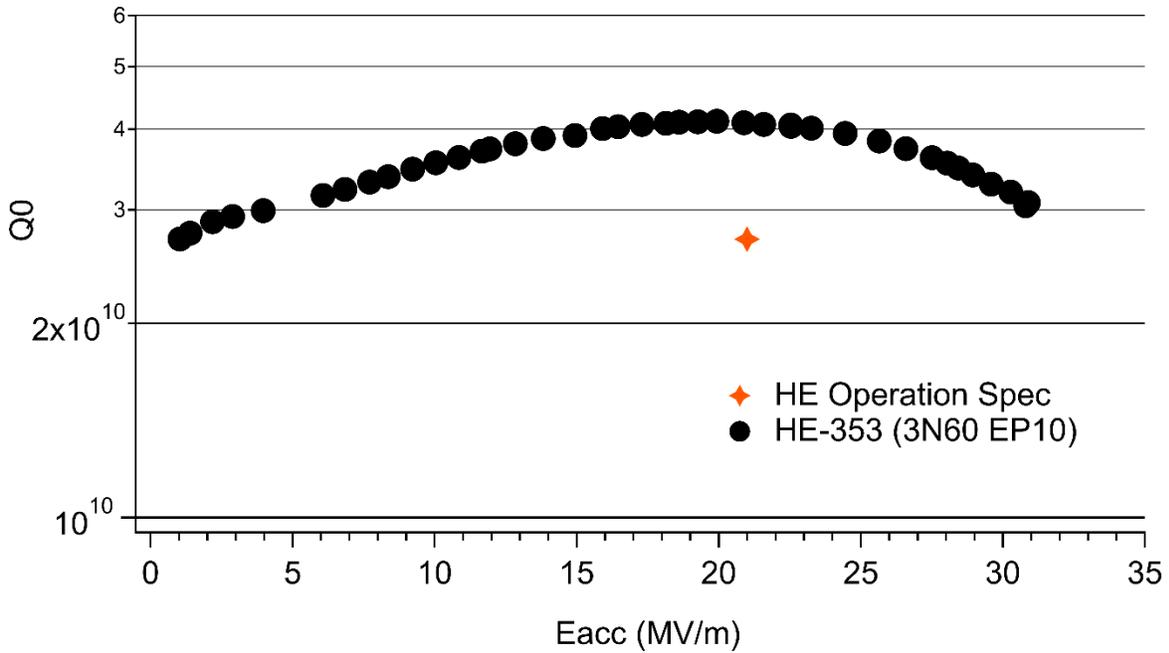

**Figure 13.** Resonance quality factor ($Q_0$) vs. effective accelerating gradient for cavity HE-353, co-processed with sample NL-147. The performance requirement of the LCLS-II HE accelerator project [ref] is noted.

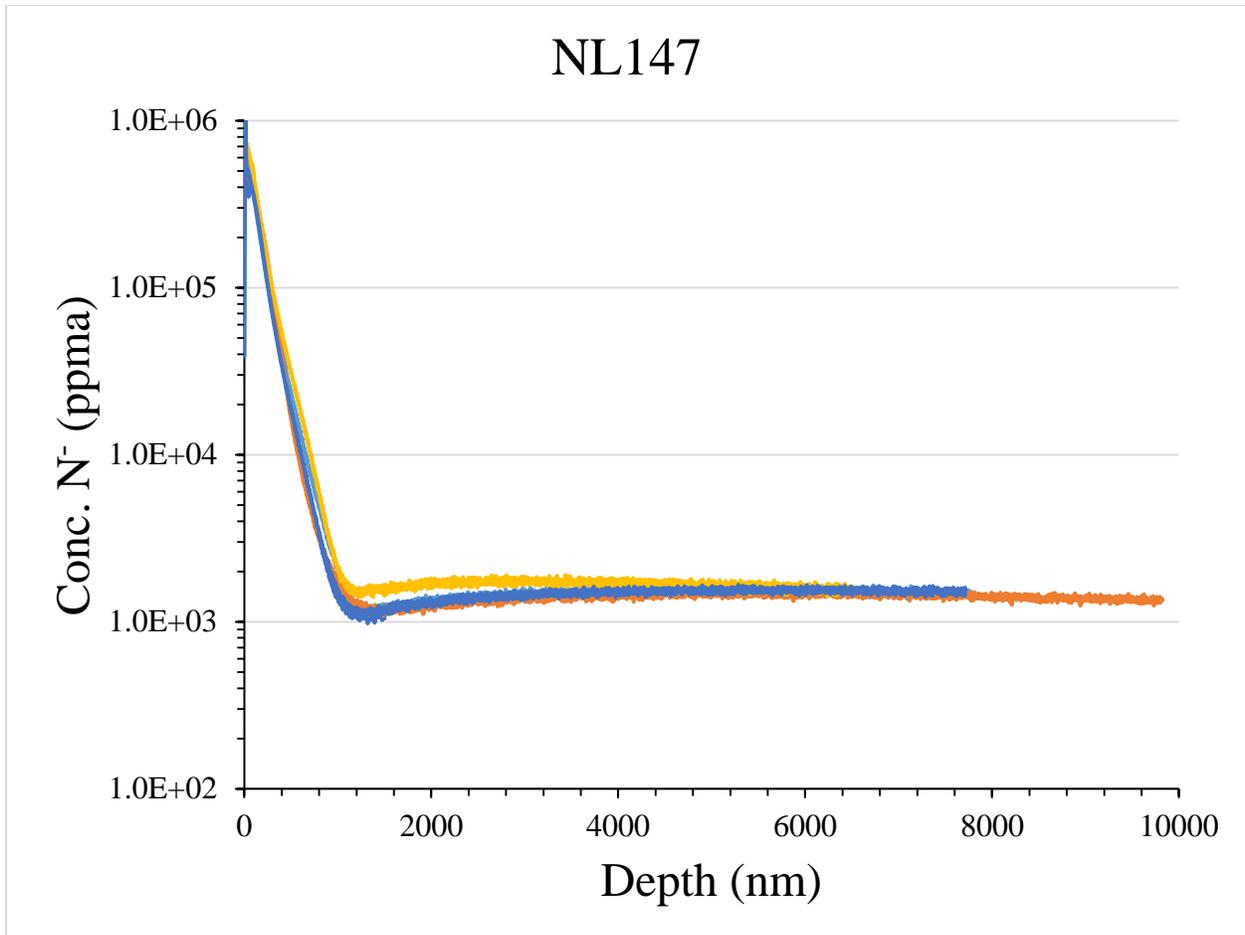

**Figure 14.** SIMS depth profiles for sample NL 147. The data show a sharp decline in NbN⁻ concentration which reaches a local minimum at the interface between the nitride layer and the bulk niobium. The concentration of the N⁻ increases in the bulk before beginning diffusion-limited decay into the bulk.

**SUMMARY AND CONCLUSIONS**

Accurate and precise quantification of nitrogen of N-doped niobium has previously proven challenging. Our previous work reported that the RSF value may deviate up to 50% by changing implant standards or simply changing the grain. This level of uncertainty is unacceptable for SRF applications and method improvement is critical.

Previous reports of nitrogen content deviation as a function of grain orientation should be examined carefully to be sure that topographic effects are adequately accounted for. Analysis of polycrystalline materials may yield different concentrations for different grains or grain combinations. This may be a result of different grain height with respect to the mass spectrometer, effectively changing the working distance. As a result, the beam may become off-centered as it passes through the secondary column, altering the baseline signal and instrument calibration. The instrument calibration obtained with use of the polycrystalline implant standard

NL 133 yielded RSF values that varied drastically from grain to grain. Furthermore, the sample holders themselves were observed to cause the working distance to deviate as a function of holder position. Additionally, the loading force to hold the samples in place was observed to cause a change in the working distance which resulted in a deviation in RSF values. All these geometrical challenges to the precise location of the sampled surface can be mitigated by centering the beam with the DTCA-x and DTCA-y apertures.

The influence of the DTCA apertures is immediately seen in the RSF variability. The RSF reproducibility was observed to have been improved from 40% to 9.6% RSD for the polycrystalline implant standard NL 133. Using a single grain implant standard, such as U52, further reduces the uncertainty to 2%. All other implant standards showed an improvement in precision when beam centering was performed. The impact of this change was further observed by analyzing the N-doped NL 115 cavity sample. Previous results were affected by surface topography and incorrectly showed a reduction of nitrogen content within the [111] grains of these samples. SEM analysis showed that the [111] grains uniquely contain a higher nitrogen content than is expected, consistent with nano-nitrides The phenomenon was reproduced and observed by analyzing NL 140 in which the presence of 'nano nitrides" were also noted. An observable difference in the average nitrogen content was noted when comparing the previous data set to the current. The lack of DTCA adjustment prior to each scan can account for some deviation. However, two separate implant standards were used for each analysis. As an absolute and traceable implant does not currently exist for nitrogen in niobium, further work is needed to refine analysis to establish excellent quantization results regardless of the implant standard used.

Improvement of the SIMS analytical methods was a critical step to performing routine analysis of SRF materials. Due to superior cavity performance associated with sample NL 147, understanding the pathway from doping to performance became a high priority. SIMS was determined to be a critical step for understanding the material science of the cavity surface preparation. The nitride layer was determined to be 1 µm followed by a minimum in nitrogen concentration at the interface. The concentration of nitrogen was observed to then increase before presumably decaying within the diffusion layer to typical bulk concentrations of ~60 ppma.

Considering all factors, we conclude that if care as described above is taken, an uncertainty of < 10 % in local nitrogen concentration with depth is reliably attainable using dynamic SIMS, for niobium materials in SRF applications.

## ACKNOWLEDGEMENTS

The authors are grateful for support from the Office of High Energy Physics, U.S. Department of Energy under grant DE-SC-0014475 to Virginia Tech. This work was co-authored by Jefferson Science Associates LLC under U.S. DOE contract DE-AC05-06OR23177. This material is based on work supported by the U.S. Department of Science, Office of Science, Office of Nuclear Physics.

## DATA AVAILABILITY

The data that support the findings of this study are available from the corresponding author upon reasonable request.

The data that support the findings of this study are available from the corresponding author upon reasonable request.